\documentclass[journal]{IEEEtran}

\ifCLASSINFOpdf
\else
\fi
\usepackage{amsmath,amssymb,amsfonts}
\usepackage{lettrine}
\usepackage{amsfonts,amssymb}
\usepackage{amsfonts}
\usepackage{url}
\usepackage{balance}
\usepackage{soul}
\usepackage{multirow}
\usepackage{colortbl} 
\usepackage{xcolor} 
\usepackage[ruled,linesnumbered]{algorithm2e}
\usepackage{algpseudocode}
\usepackage{amsmath}
\usepackage{graphics}
\usepackage{epsfig}
\usepackage{color}
\usepackage[colorlinks,
            linkcolor=blue,
            anchorcolor=blue,
            citecolor=blue,
            urlcolor=blue]{hyperref}
\usepackage{float}
\usepackage{verbatim}
\usepackage{cite}
\usepackage{orcidlink}
\usepackage{multirow}
\usepackage{color}
\usepackage{colortbl,hhline}
\usepackage{xcolor}
\usepackage{colortbl}
\usepackage{nomencl}
\makenomenclature
\usepackage{soul}
\usepackage{array}
\usepackage{diagbox}
\usepackage{arydshln}
\usepackage[T1]{fontenc}
\usepackage{booktabs}
\usepackage{makecell} 
\usepackage{booktabs} 
\usepackage{multirow}
\usepackage[normalem]{ulem}

\useunder{\uline}{\ul}{}
\begin{document}

\title{A Novel Coronary Artery Registration Method Based on Super-pixel Particle Swarm Optimization}

\author{Peng Qi, Wenxi Qu, Tianliang Yao, Haonan Ma, Dylan Wintle,\\
        Yinyi Lai, Giorgos Papanastasiou, Chengjia Wang} 
%

\markboth{Preprint}%
{Yao \MakeLowercase{\textit{et al.}}: Multi-Agent}

\maketitle

\begin{abstract}
Percutaneous Coronary Intervention (PCI) is a minimally invasive procedure that improves coronary blood flow and treats coronary artery disease. Although PCI typically requires 2D X-ray angiography (XRA) to guide catheter placement at real-time, computed tomography angiography (CTA) may substantially improve PCI by providing precise information of 3D vascular anatomy and status. To leverage real-time XRA and detailed 3D CTA anatomy for PCI, accurate multimodal image registration of XRA and CTA is required, to guide the procedure and avoid complications. This is a challenging process as it requires registration of images from different geometrical modalities (2D -> 3D and vice versa), with variations in contrast and noise levels.  In this paper, we propose a novel multimodal coronary artery image registration method based on a swarm optimization algorithm, which effectively addresses challenges such as large deformations, low contrast, and noise across these imaging modalities. Our algorithm consists of two main modules: 1) preprocessing of XRA and CTA images separately, and 2) a registration module based on feature extraction using the Steger and Superpixel Particle Swarm Optimization algorithms. Our technique was evaluated on a pilot dataset of 28 pairs of XRA and CTA images from 10 patients who underwent PCI. The algorithm was compared with four state-of-the-art (SOTA) methods in terms of registration accuracy, robustness, and efficiency. Our method outperformed the selected SOTA baselines in all aspects. Experimental results demonstrate the significant effectiveness of our algorithm, surpassing the previous benchmarks and proposes a novel clinical approach that can potentially have merit for improving patient outcomes in coronary artery disease.
\end{abstract}


\section{Introduction}
Percutaneous Coronary Intervention (PCI) is a minimally invasive technique that can restore the blood flow of blocked coronary arteries and treat coronary artery disease \cite{r1}. During PCI procedures, X-ray angiography (XRA) is used to obtain real-time 2D projection images that guide catheter placement with high precision \cite{r2}. However, the projective nature of XRA lacks depth information, making the interpretation of complex 3D vascular anatomy challenging. In contrast, preoperative computed tomography angiography (CTA) can provide 3D visualization of coronary vessels but suffers from the disadvantages of radiation exposure and the inability for real-time imaging.


To gain precise information about the disease status and vascular anatomy, recent research has focused on the registration of intraoperative XRA images with preoperative CTA volumes. The goal is to combine their complementary advantages to facilitate navigation through occluded lesions and twisted vascular anatomies during PCI procedures. Intraoperative XRA provides real-time imaging and tool guidance, while preoperative CTA offers 3D context for enhanced visualization and navigation. However, several challenges need to be overcome in this 3D/2D registration task. The dimensional disparities between 3D volumes and 2D images necessitate sophisticated projection models to enable comparison and alignment. Secondly, the coronary arteries undergo complex nonlinear deformations caused by cardiac motion and respiration between the preoperative CTA and intraoperative XRA acquisitions. Robust optimization schemes must be developed to effectively handle these challenges. Therefore, the registration of 3D/2D images remains an open challenge and an active research area for enhancing PCI procedures.

In this study, we propose a novel method for registering CTA and XRA images of coronary arteries based on Superpixel Particle Swarm Optimization (SPSO). The method consists of the following modules: keypoint and edge extraction of the DSA image, keypoint extraction of the CTA image, and vascular matching based on particle swarm optimization. In the keypoint and edge extraction module of the DSA image, an improved deep learning network is used to extract the XRA structural features. Layer peeling, distance transformation and sub-pixel corner detection is employed to extract the 2D vascular edge and cross-point features. For the keypoint extraction module of the CTA image, mimics segmentation is applied to extract the 3D vascular edge and cross-point features. In the vascular matching module based on particle swarm optimization, our method utilizes a similarity matching approach using particle swarm optimization search. This approach enables the feature point matching algorithm and optimization search algorithm for the feature points of XRA and CTA images, achieving high-precision and high-efficiency 3D-2D vascular registration. Overall, our proposed method offers a robust and accurate approach for registering CTA and XRA images of the coronary artery, facilitating precise and efficient 3D-2D vascular registration.

The main contributions of this paper can be summarized as follows:

\begin{itemize}
    \item A new particle swarm optimization approach has been introduced for 3D-2D registration of XRA and CTA images. This method offers rapid convergence, minimal parameter configuration, and straightforward implementation.
    
    \item •	We designed an "extraction and discretization" procedure for efficient extraction of vessel centerlines, with which the accuracy of 2D-3D object matching is then done by our simple weighted closest point method. extraction and discretization of vessel centerlines, weighted matching, which can improve the accuracy of the matching.
    
    \item •	A deep learning-based vessel segmentation technique for DSA images is proposed, aiming to achieve more accurate segmentation results and improve registration precision.
\end{itemize} 

\begin{figure*}[!ht]
\centerline{\includegraphics[width=1.0\textwidth]{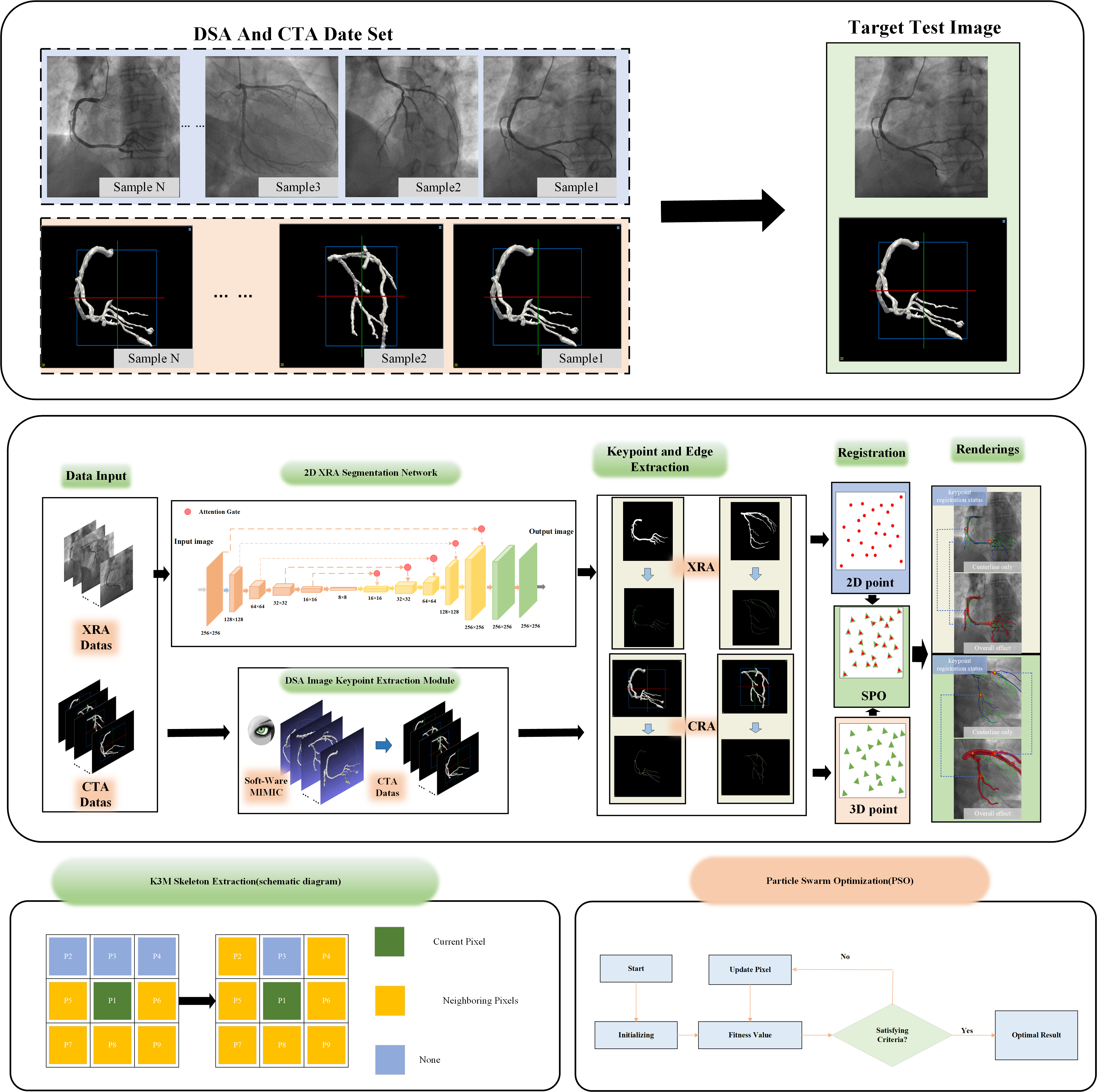}}
\caption{Workflow of the Coronary Artery Registration}
\end{figure*}

\section{Related Works}
In this section, 3D/2D registration in coronary angiographic image sequences provides useful structural information for diagnosis and treatment guidance of cardiovascular diseases. Existing methods for 3D/2D registration can be broadly categorized into two classes: intensity-based approaches and feature-based approaches.

\subsection{Approach Based on intensity}

Intensity-based approaches utilize the gray intensity information in the image sequences for vascular tracking. Sundar et al. \cite{r3} used the cumulative phase shift in the spectral domain of consecutive frames. Panayiotou et al. \cite{r4} applied manifold embedding to reduce image dimension. Fischer et al. \cite{r5} developed an isometric feature mapping method to map each image to a 1D signal. Ma et al. \cite{r6} applied principal component analysis on angiograms. These methods are all based on analyzing the global gray intensity variations in the sequences to extract signals representing motion patterns. However, the motion patterns are complex with mixed effects of respiratory and cardiac motions. Using pure intensity information makes it difficult to separate respiratory motion from other motions.

\subsection{Approach Based on Structural Features}
Feature-based approaches rely on extracting and tracking vascular features such as bifurcations, centerlines, or landmarks. Sun and Yu \cite{r7} improved the snake model to track and deform open curves for vascular tracking. Zhang et al. \cite{r8} tracked landmarks on vessels and fitted target points using a B-spline to obtain vascular structures. Shin et al. \cite{r9} used image registration to determine possible vascular locations and a Markov model to optimize the positions. Cheng et al. \cite{r10} viewed vessels as multi-target points and used low-rank tensor approximation to track curvilinear structures. Chu et al. \cite{r11} utilized deep learning to extract vascular features and employed multi-target tracking to obtain vascular structures. Feature-based methods can achieve higher accuracy by utilizing local structural constraints. However, they still face challenges in accurately establishing point correspondences due to similar vascular textures and complex backgrounds. The error propagation in sequential tracking also affects the robustness. In addition, the topological connection relations between tracked key points are difficult to determine in these methods.

In summary, intensity-based methods have difficulties in separating respiratory motion from other complex patterns, while feature-based methods lack robustness due to inaccuracies in feature extraction and correspondence establishment. Hybrid methods are also explored to exploit both intensity and feature information. Vermandel et al. \cite{r12} registered the segmentations of vasculature using intensity-based methods. Mitrović et al. \cite{r13} matched the projected vessel orientations with 2D gradients for registration. However, challenges remain in handling complex motion patterns and avoiding error propagation.

In this work, we propose a novel method for processing XRA and CTA images, which consists of two major steps: segmentation and discretization of the vessels and cross points, and alignment of 2D and 3D super-pixels using particle swarm optimization (PSO) algorithm. A super-pixel is defined as a discrete point with edge and point features. The proposed method aims to achieve the matching of 2D and 3D vascular images by minimizing the distance between the corresponding super-pixels. Our method overcomes the problem of many parameters in general deep learning methods for 2D and 3D image registration and leverages the advantages of PSO, such as fast convergence speed, low parameter usage, and simple implementation.

\section{Methodology}

In this section, we present a novel alignment approach, tailored for aligning XRA and CTA images of coronary arteries. The workflow of our method is shown in Fig 1. 

The overall framework of the proposed 3D-2D vascular matching method based on super-voxel searching is presented in Figure 1. This framework consists of modules for DSA image keypoint and edge extraction, CTA image keypoint extraction, and vascular matching based on particle swarm optimization.

In the DSA image keypoint and edge extraction module, we obtain the 2D vascular structure through vessel segmentation using residual attention learning. Centerline extraction is performed by utilizing layer peeling and first and second-order grayscale processing of sub-pixel corners. Building upon the 2D vascular structure, we acquire the 2D vascular skeleton using distance transformation-based moments, resulting in the generation of 2D vascular super-voxels.

Within the CTA image keypoint extraction module, we obtain the 3D vascular structure and its centerlines using dynamic threshold segmentation. Applying a pose transformation (3D to 2D projection) to this structure generates the projected 2D vascular super-voxels.
Within the vascular matching module based on particle swarm optimization, a similarity match is performed using weighted super-voxels obtained from the aforementioned modules: 2D vascular super-voxels and projected 2D vascular super-voxels.

Finally, our proposed method offers an effective approach for registering XRA and CTA images of the coronary artery, leveraging super-voxel searching and particle swarm optimization to achieve accurate 3D-2D vascular matching

When the matching outcome meets the requirements, the final result is achieved. The following subsection will provide a detailed description of these modules.

\subsection{DSA Keypoint and Edge Extraction Module}

To establish common features and connections between 2D and 3D vasculatures, keypoint and edge extraction are performed on DSA images. Vessel segmentation on DSA images is achieved using a deep-learning segmentation model based on UNet. The architecture of its network is shown in Fig 2.

Vessel segmentation of DSA images is performed using the UNet deep learning model \cite{r14}. UNet is a commonly used semantic segmentation model that is well-suited for medical image segmentation tasks. The UNet model consists of an encoder and a decoder, enabling effective extraction of semantic features from the image and restoration of spatial resolution. The encoder utilizes the Conv class as the basic convolutional neural network module, serving as the backbone network for extracting high-level semantic features from the input image. These features undergo a series of downsampling operations, gradually reducing the size and number of channels in the feature maps. The decoder adopts the structure of UNet, restoring the spatial resolution through upsampling and feature concatenation with the encoder. These feature concatenations form skip connections, which help propagate richer spatial information through the network. The modules typically employ the CBAM architecture (channel attention and spatial attention) to learn important features across channels and spatial positions. The output of the network is passed through a 1x1 convolution and a sigmoid function to obtain the vessel segmentation results. The binary cross-entropy loss function is employed to simultaneously optimize the segmentation performance for both foreground and background.

\begin{figure}[ht]
\centerline{\includegraphics[scale=1.1]{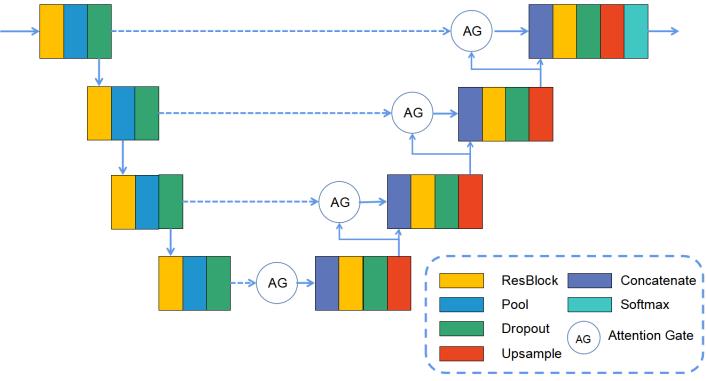}}
\caption{Illustration of the Segmentation Network Structure}
\end{figure}

After segmenting the vessel contours from XRA images, we further extract the vessel centerlines and skeletons using layer peeling and distance transformation. Layer peeling is a technique that progressively removes layers from the vessel segmentation result and selects the central gray value. For skeletonization, we adopt the K3M skeleton erosion algorithm \cite{r15}, which is an iterative erosion boundary-based image thinning method. The goal of this algorithm is to gradually reduce the object boundaries in a binary image until only a single-pixel-wide skeleton remains. The algorithm schematic is shown in Figure 3, The mathematical formulation of this algorithm is as follows:

Let $X$ be a binary image and $B$ be a structuring element. We denote by $B+X$ the subset of $X$ that matches $B$, and by $B-X$ the subset of $X$ that is complementary to $B$. The algorithm can be expressed as follows:

\begin{equation}
X^{(k+1)} = X^{(k)} - \bigcup_{i=1}^4 (B_i + X^{(k)}) \cap (B_i - X^{(k)}) 
\end{equation}
where $X^{(0)}=X$, $X^{(k+1)}$ is the thinned image after the $(k+1)^{th}$ iteration and $B_{i}$ are four structuring elements defined as:
\begin{equation}
B_1 = \begin{bmatrix}
0 & 0 & 0 \\
0 & 1 & 0 \\
1 & 1 & 1
\end{bmatrix}, 
\\
B_2 = \begin{bmatrix}
0 & 0 & 0 \\
1 & 1 & 0 \\
1 & 1 & 0
\end{bmatrix}, 
\\
B_3 = B_1^T, 
B_4 = B_2^T
\end{equation}

The algorithm stops when no more pixels can be removed, i.e., $X^{(k+1)}=X^{(k)}$. The final result is a one-pixel-wide skeleton of the original image.

\begin{figure}[ht]
\centerline{\includegraphics[scale=0.45]{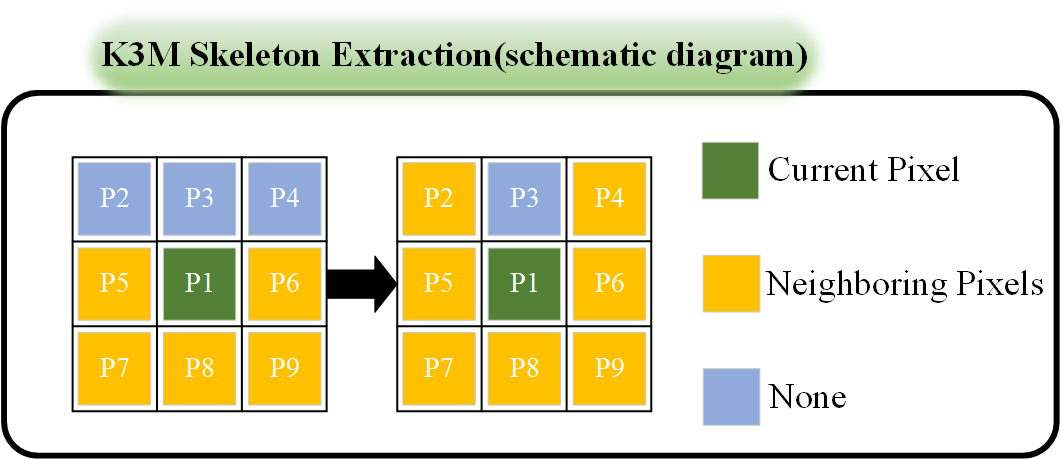}}
\caption{K3M Skeleton Extraction(schematic diagram)}
\end{figure}

To precisely extract key points and edges from XRA vascular images, further sub-pixel corner detection is necessitated. We refer to the Shi-Tomasi corner detection algorithm and gray moment model for conjoint sub-pixel corner extraction. However, this leads to an overabundance of key points, including some redundant loci within the vessels, which necessitate filtration.

The Shi-Tomasi corner detection algorithm is an improved formulation of the Harris corner detection algorithm, designed to identify regions in an image exhibiting robust contrast and directional variances. Fundamentally, if the gray values of an image locale undergo significant alterations in any orientation due to movement, then the locale can be classified as a corner. The Shi-Tomasi algorithm employs a matrix M to characterize the gray variabilities in an image region, wherein the elements of M are the products of the gradient valuations (partial derivatives of gray values). The two eigenvalues $\lambda_{1}$ and $\lambda_{2}$ of $M$ represent the extent of gray variation in two orthogonal orientations of the image region. According to the Shi-Tomasi algorithm, if both eigenvalues of $M$ surpass a definite threshold, the region is a corner. The mathematical description for the Shi-Tomasi algorithm is as follows:

Let $X$ be a grayscale image, B be a structuring element, $B+X$ represent the subset of $X$ that is identical to $B$, and $B-X$ represents the subset of X that is complementary to $B$. The Shi-Tomasi algorithm can be represented as:

\begin{equation}
M=\sum\limits_{(x,y)}w(x,y)\begin{bmatrix}l_{x}^{2}&l_{x}l_{y}\\l_{x}l_{y}&l_{y}^{2}\end{bmatrix}
\end{equation}

Thereby accomplishing the discretization of the centrelines of XRA vascular architectures.

First, we perform binary thresholding on the XRA image to separate the vascular structures from the background. We choose an appropriate threshold $T$ such that pixels with grayscale values greater than $T$ are set to 1, representing vessels, while pixels with grayscale values less than or equal to $T$ are set to 0, representing the background.

Second, we calculate the original moments $M_{pq}$ and the central moments $\mu _{pq}$ of the binary image using the following equations:

\begin{equation}
M_{pq}=\sum\limits_{x=0}^{M-1}\sum\limits_{y=0}^{N-1}x^{p}y^{q}f(x,y)
\end{equation}

Where M and N are the image dimensions,$ (x,y)$ are the pixel coordinates, p and q are the order of the moment, and $f(x,y)$ is the binary pixel value (0 or 1) at coordinates $(x,y)$.

\begin{equation}
\mu_{pq}=\sum\limits_{x=0}^{M-1}\sum\limits_{y=0}^{N-1}(x-\bar{x})^{p}(y-\bar{y})^{q}f(x,y)
\end{equation}

Where $\bar{x}$ and $\bar{y}$ are the centroid coordinates defined as:

\begin{equation}
\bar{x}=\frac{M_{10}}{M_{00}}
\end{equation}
\begin{equation}
\bar{y}=\frac{M_{01}}{M_{00}}
\end{equation}

Third, we use the central moments $\mu_{pq}$ to calculate the centroid $(x_c, y_c)$of each connected vascular region as:

\begin{equation}
x_{c}=\bar{x}+\frac{\mu_{10}}{\mu_{00}}
\end{equation}
\begin{equation}
y_{c}=\bar{y}+\frac{\mu_{01}}{\mu_{00}}
\end{equation}

Where $\mu_{00}$ is the area of the vascular region. The centroids $(x_c, y_c)$ represent the bifurcation points of the vasculature. By extracting these key points along with the vascular boundaries, we can reconstruct the complete vascular structures from the XRA image. The entire process is depicted in Figure 4.

\begin{figure}[ht]
\centerline{\includegraphics[scale=0.32]{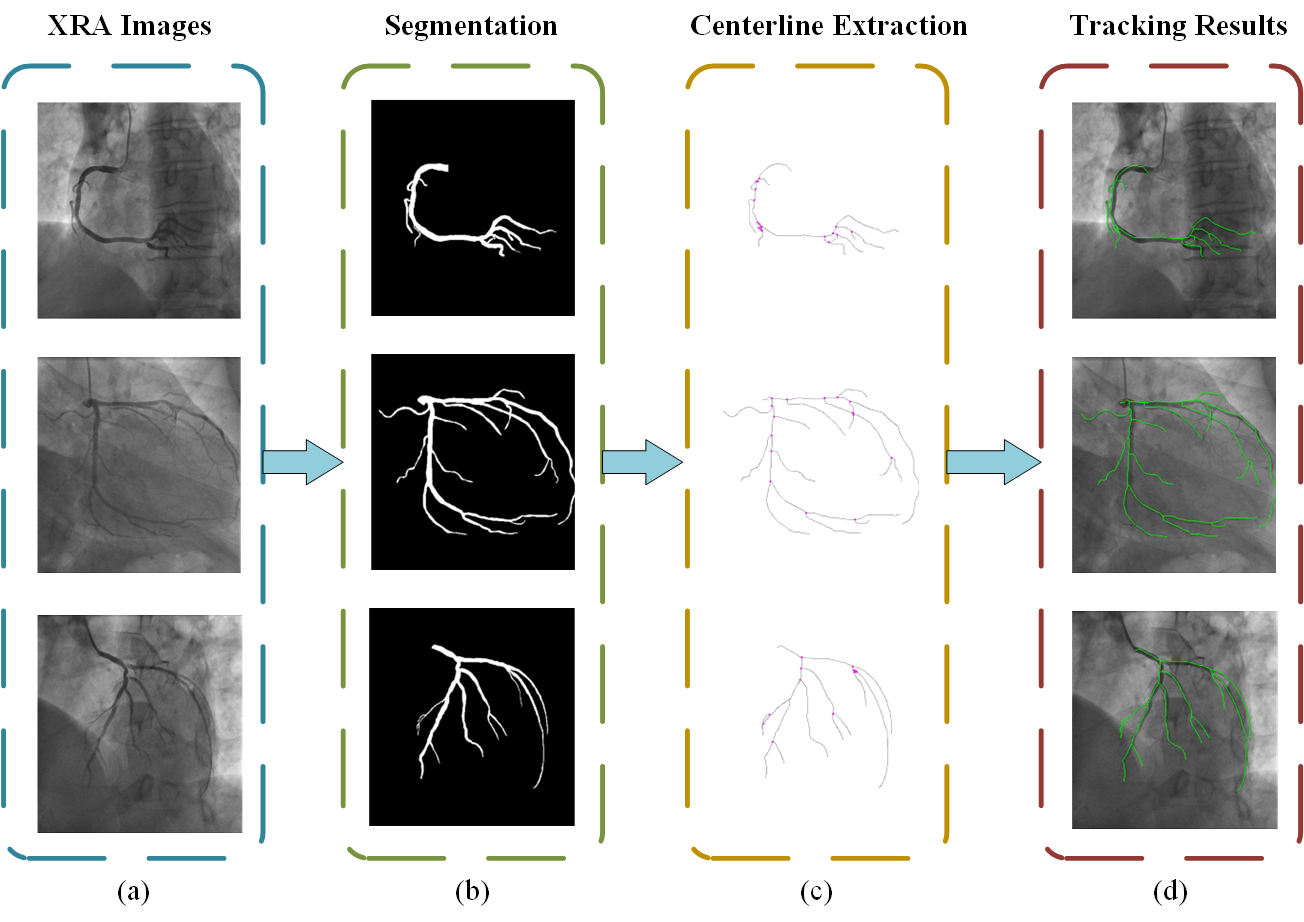}}
\caption{Visualization of the Registration Process}
\end{figure}

\subsection{CTA Image Keypoint Extraction Module}
In the field of medical image analysis, the extraction of key points and features plays a vital role in various applications. This research focuses on the development of a DSA (Digital Subtraction Angiography) Image Keypoint Extraction Module that aims to obtain shared features between DSA and CTA (Computed Tomography Angiography) modalities.DSA images provide valuable information about the vascular structures in the human body. However, to gain a comprehensive understanding and perform comparative analysis, it is essential to process CTA images and extract key points that can be compared with the DSA images. This paper proposes a method that utilizes dynamic threshold segmentation and \href{www.materialise.com}{MIMICS}\cite{r16} software for the extraction of key points from CTA images.\cite{r17}

The proposed method employs dynamic threshold segmentation, a technique widely used in medical image processing. The MIMICS software, specifically designed for CTA images, is utilized to carry out the segmentation process. By applying dynamic threshold segmentation, 3D vascular structures and their corresponding centerlines are extracted from the CTA images.

The application of the proposed method yields effective extraction of 3D key points from CTA images. These key points serve as common features that can be compared with the key points and edges extracted from DSA images. The identification of common features facilitates further analysis and comparison between the DSA and CTA modalities, contributing to a comprehensive understanding of the vascular structures.

The DSA Image Keypoint Extraction Module presented in this research enables the extraction of key points from CTA images using dynamic threshold segmentation and the MIMICS software. By obtaining shared features between DSA and CTA modalities, this module supports the analysis and comparison of vascular structures. Further research and development in this area are expected to enhance medical image analysis and contribute to improved diagnosis.

\subsection{Particle Swarm Optimization for 3D-2D Vascular Matching}

After obtaining the 3D vascular structure and its centerline from the CTA image, we need to project the 3D vasculature onto the 2D plane to obtain the projected 2D vascular super-voxels, and then find the optimal matches between the 3D vasculature and the corresponding positions in the XRA image. To achieve this, we perform a similarity matching based on particle swarm optimization (PSO)\cite{r18} between the discretized 2D vascular points and the projected 2D vascular points.The corresponding algorithm framework can be seen in Figure 5.

\begin{figure}[ht]
\centerline{\includegraphics[scale=0.50]{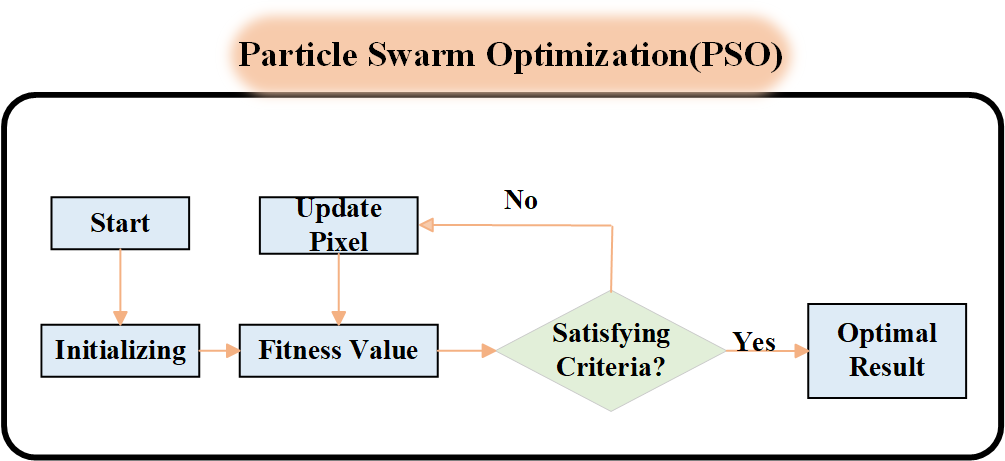}}
\caption{Particle Swarm Optimization(PSO)}
\end{figure}

For this purpose, we adjust the six degrees of freedom of the virtual camera (consisting of three rotation angles and one translation vector). After each adjustment of the virtual camera’s pose, we project the 3D super-voxels onto the 2D vascular plane, generating the projected 2D vascular super-voxels. Then, we conduct a weighted similarity matching between the 2D vascular super-voxels (XRA) and the projected 2D vascular super-voxels (CTA) using PSO.

For each discrete point, we use a velocity update formula to guide how each particle adjusts its movement velocity in each iteration, aiming to move in a more optimal direction. The velocity update formula is as follows:

\begin{equation}
    v_{i}^{t+1}=\omega v_{i}^{t}+c_{1}r_{1}(p_{i}^{t}-x_{i}^{t})+c_{2}r_{2}(g^{t}-x_{i}^{t})
\end{equation}

Where $v_{i}^{t}$ represents the velocity vector of the i-th particle at the t-th iteration. $x_{i}^{t}$ represents the position vector of the i-th particle at the t-th iteration.
$p_{i}^{t}$ represents the individual best position vector of the i-th particle at the t-th iteration. $g^{t}$ represents the global best position vector at the t-th iteration. $\omega$ is the inertia weight coefficient. $c_{1}$ and $c_{2}$ are the cognitive and social learning factors, respectively. $r_{1}$ and $r_{2}$ are random numbers in $[0,1]$.

The position update formula describes how each particle changes its position based on its velocity in each iteration, enabling exploration of a wider search space. The position update formula is as follows:

\begin{equation}
    x_{i}^{t+1}=x_{i}^{t}+v_{i}^{t+1}
\end{equation}

The weighted similarity measure is computed and compared to a threshold value to determine if it meets the desired criteria. If the current weighted similarity measure is not greater than the threshold, the process continues by adjusting the pose of the virtual camera, projecting the vasculature, and recalculating the weighted similarity measure. The iteration termination conditions are based on the final result's weighted similarity measure, specifically using the Normalized Cross-Correlation (NCC). To establish these conditions, a user-defined parameter "e" is set, which is determined through experimental exploration and practical requirements. If the NCC value exceeds "e," it indicates that the desired matching effect has been achieved, or the maximum iteration count has been reached, or the cost function has converged.

If the NCC value is not greater than "e," it signifies the need to return to the pose transformation applied during the extraction of the 3D vascular structure and centerline. New projected 2D vascular points are obtained for matching, and the iterative process continues until the matching result satisfies the desired NCC threshold, reaches the maximum iteration count, or converges in terms of the cost function. Once these conditions are met, the 3D-2D matching result can be output.

\subsection{Evaluation Metrics}
Evaluation of semantic segmentation requires quantitative metrics to measure the segmentation quality. The Dice coefficient is a commonly used method to measure the similarity between the segmented regions. The Dice coefficient is calculated as:

\begin{equation}
Dice(A,B) = \frac{2|A\cap B|}{|A| + |B|}
\end{equation}

Where $A$ and $B$ are the segmented regions, $|A\cap B|$ is the number of pixels in the intersection of A and B, and $|A|$ and $|B|$ are the total numbers of pixels in $A$ and $B$, respectively.

To evaluate the performance of the proposed registration method, we use the minimum projected distance (mPD) \cite{r19} as a metric. The mPD is defined as follows: for each point in the 3D vascular structure, we project it onto the 2D vascular plane along all possible directions and choose the direction that minimizes the projection distance. The mPD is the average of the minimum projection distances for all points in the 3D vascular structure. The mPD can better reflect the quality of the registration result on the 2D image because it considers the impact of different projection directions on the registration error. We calculate the mPD using the following formula:

\begin{equation}
mPD(A,B)=\frac{1}{|A|}\Sigma_{a \in A}d(a,P_{B}(a))
\end{equation}

Where $A$ is the set of points in the 3D vascular structure, $B$ is the set of points in the 2D vascular plane, $|A|$ is the cardinality of A, $d(a, P_{B}(a))$ is the minimum projection distance for point $a\in A$, and $P_{B}(a)$ is the projection of point a onto plane $B$.

The mPD metric is particularly valuable in assessing registration performance on 2D images as it takes into account the impact of various projection directions, providing a comprehensive evaluation of the registration accuracy. By considering the minimum projection distance, mPD offers a more nuanced and informative assessment compared to traditional error measures.

The mPD metric serves as a robust evaluation criterion, allowing for a comprehensive assessment of the registration quality by considering the projection distances in various directions. Its application facilitates a more insightful understanding of the accuracy and reliability of the registration method on 2D images.

\section{Experiments}
\subsection{Datasets}
This dataset contains CT and XRA images of 28 patients retrospectively at the Department of Cardiology, Tongji Hospital of Tongji University. The CTA images were acquired on the Siemens SOMATOM Force CT platform. The slice number ranged from 196 to 673. The resolution of each slice was 512 × 512 with 0.35 × 0.35 × 0.75 (unit: mm) voxel size correspondingly, and ECG-gating.

The XRA sequences were acquired on Phillip Azurion 7M12, with a frame rate of 15 fps. The XRA sequences were restricted to contain at least one complete cardiac cycle, ranging from 20 to 50 approximately. Each frame had a 512 × 512 resolution with 0.38 × 0.38 (unit: mm) pixel size. Furthermore, only contrast agent-filled frames in each sequence are available to match the corresponding CT images because these frames show the complete coronary artery structures, excluding similar images.

To train the segmentation model, 1571 XRA images annotated with semantic segmentation labels are utilized in the dataset. They are randomly split into training, validation, and test sets at a ratio of 8:1:1. Specifically, 1257 images $(80\%)$ were assigned to the training set. 157 images $(10\%)$ formed the validation set. The remaining 157 images $(10\%)$ made up the test set. To prevent data leakage, no images were shared between the splits.

\subsection{Implementation Details}

Based on the CMake build tool, the software environment is: Visual Studio 2017 compiled release version, the operating system is Windows 10 Home Chinese Edition, hardware environment is a 16-core 2.3GHZ 11th Gen Intel® Core™ i7-11800H processor and NVIDIA GeForce RTX 3070 Laptop GPU. 

Based on the above experimental settings, the number of cycles is generally limited to 1000. If you need to increase the time, you can reduce unnecessary cycles. Considering that a coronary artery sample usually needs to be segmented into multiple vessel segments for a complete extraction, it takes about 0.04 seconds. 

Since the data is generated based on the known point correspondence relationship, mPD is used in the loss function instead of soft dtw when performing experiments on the dataset.

\begin{table*}[ht]
\caption{Parameters of 3D-2D Registration Methods}
\begin{center}
\resizebox{\textwidth}{!}{
\begin{tabular}{cccc}
\hline
  & \makecell{\textbf{Stop Condition} \\ The Maximum Number of Iterations} & \textbf{Optimizer} & \textbf{\makecell{Deformation Model Parameter \\ Calculation Method}}\\
\hline
ICP-BP  & 300 & Levenberg-Marquard & 3D-2D fitting  \\
ICP-PnP & 300 & PnP & 3D-2D fitting  \\
DT & 300 & Levenberg-Marquard & optimization \\
CS & 10 & Quasi-Newton & 3D-2D fitting  \\
Tree & 10 & Quasi-Newton & 3D-2D fitting \\
GMM & 100 & Quasi-Newton & optimization  \\
OGMM & 100 & Quasi-Newton & optimization \\
PSO-EM & 500 & PSO & optimization  \\
GRad & 100 & Quasi-Newton & optimization \\
\textbf{SPSO(Ours)} & 400 & Swarm Optimization & optimization  \\
\hline
\end{tabular}
\label{tab1}
}
\end{center}
\end{table*}

\subsection{Registration Experiment 1}
In this section, we evaluate our proposed 3D-2D registration method on the dataset and compare its performance with nine competing methods.

These methods can be categorized into four groups based on their main techniques: ICP-based methods, topology-based methods, Gaussian mixture model-based methods, and feature point-based methods. The ICP-based methods include the ICP extension method based on back projection (ICP-BP) and the ICP extension method using a closed-form solution of the PnP problem to compute the transformation matrix (ICP-PnP). The topology-based methods include the tree topology alignment method based on vascular topology consistency (Tree)\cite{r20} and the method using the Distance Transform of a pre-computed segmented vessel tree to speed up the ICP strategy (DT) \cite{r21}and method based on a paired frame of Curvilinear Structure (CS) \cite{r22}. The Gaussian mixture model-based methods include the 3D/2D Gaussian mixture model method (GMM) \cite{r23} and the orientation-weighted Gaussian mixture model (OGMM)\cite{r24}. The feature point-based methods include the 3D/2D feature point alignment method based on expectation maximization framework (PSO-EM) \cite{r25} and the vascular 3D/2D rigid alignment method based on 3D vessel centerline and 2D image gradient (Grad) \cite{r26}. We evaluated our proposed methods using several parameter configurations. Table 1 summarizes the key settings for each approach. The Grad method relies solely on 2D image gradients, therefore we only tested it on real data. For the GMM and OGMM models, the position Gaussian scale was set to 1 pixel to model local deformations. The orientation Gaussian scale in OGMM is .02 radians, and since the Grad method uses the gradient of the two-dimensional image, this method is only tested on real data. The parameter settings, optimizer selection, and deformation model parameter calculation methods of the above methods are shown in Table 1.

The experiment was conducted to compare the proposed method with manually annotated XRA segmentation images by medical experts. Figure A shows the basic registration results of our method, including the central line registration image, three-dimensional model, and two-dimensional image registration image, demonstrating a visually favorable registration effect. The result can be seen in Figure 6.

\begin{figure}[ht]
\centerline{\includegraphics[scale=0.32]{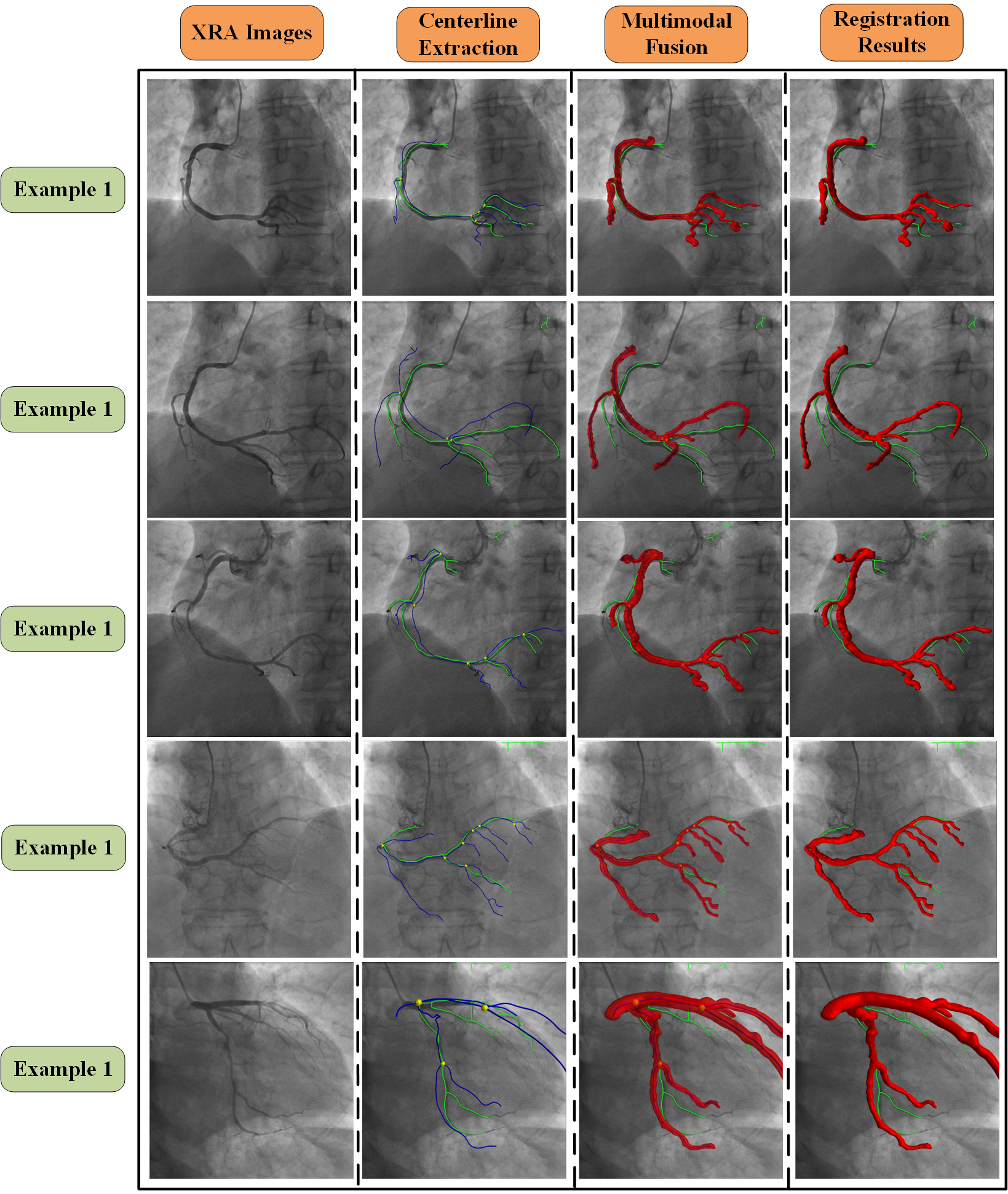}}
\caption{Image registration result flow}
\end{figure}

Figure 7 displays the registration results of our method specifically for the left and right coronary arteries. 

\begin{figure}[ht]
\centerline{\includegraphics[scale=0.4]{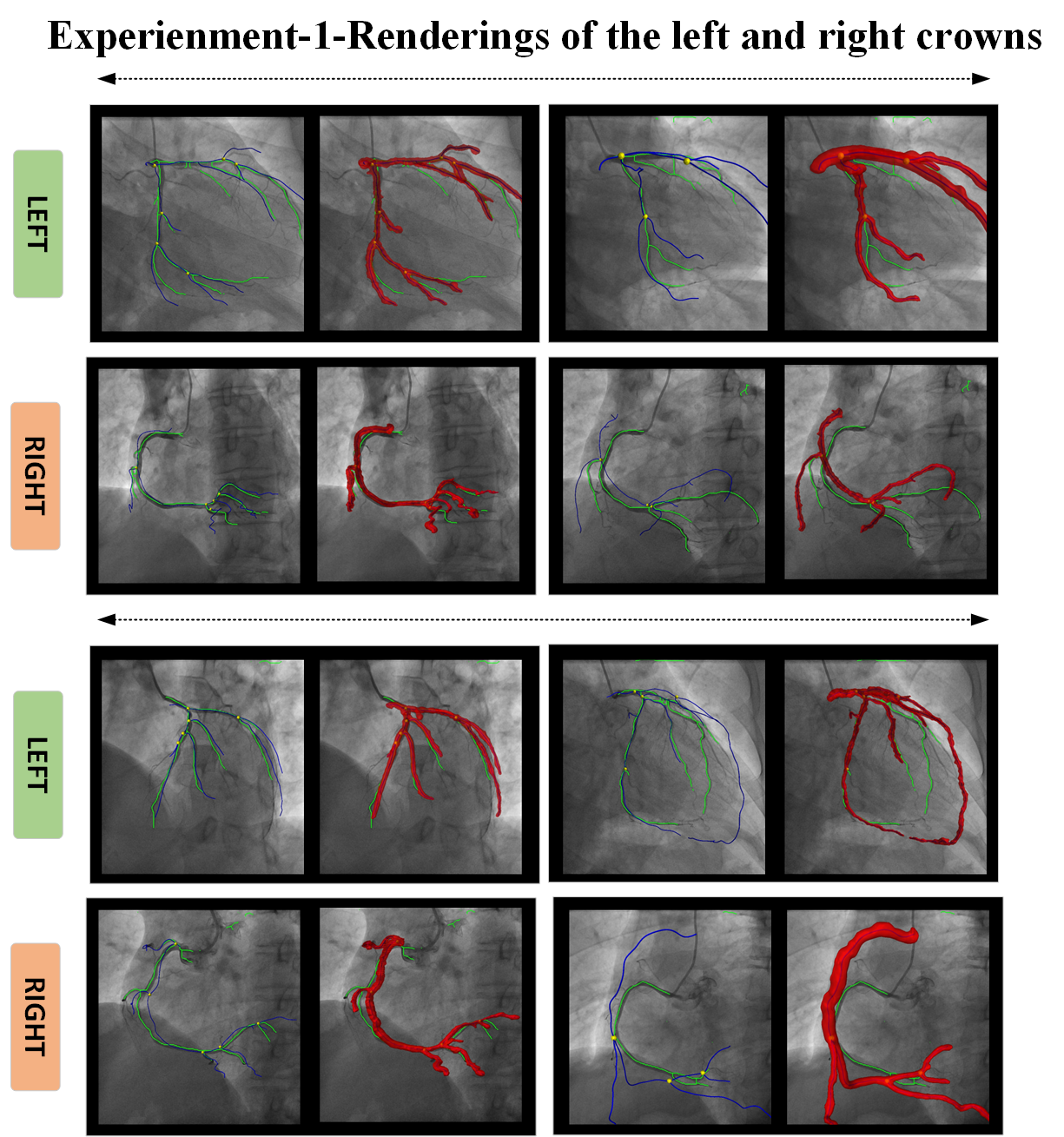}}
\caption{ registration results of L and R coronary arteries}
\end{figure}

\subsection{Registration Experiment 2}

In order to evaluate the impact of different deep learning segmentation models on the registration accuracy of DSA images, we further conducted performance testing on the dataset using different trained models for image segmentation. We compared the performance of six competitive methods: 1) R2Net, 2) ResUNet++, 3) TransResUNet, 4) U2Net, 5) UNet, and 6) VGGNet. We trained all models for 200 epochs on a dataset of 968 pairs of DSA images, and the training process was recorded using checkpoints. The best-performing model based on the test results was selected for comparison experiments, ensuring that all models had converged throughout the process.
Using the selected best-performing model, we performed segmentation on a test dataset of 60 pairs of DSA images and obtained segmentation performance metrics. For the registration experiment, the DSA images segmented using different models were processed using the DSA keypoint and edge extraction module. The corresponding CT data was processed using the CTA vessel keypoint extraction module. Finally, the particle swarm optimization 3D-2D vessel matching module was used for registration. Each model and its corresponding CT data underwent 10 registrations, and the performance metrics were obtained using the average method and by removing outliers. In other words, each method underwent at least 600 registrations. The registration results of the six competitive methods are shown in Figure XXX, and the performance and registration results of the six methods are presented in Table 2.

\begin{table*}[ht]
\caption{Parameters of 3D-2D Registration Methods}
\begin{center}
\centering
\resizebox{\textwidth}{!}{
\begin{tabular}{c|ccc|cccc}
\hline
 & \multicolumn{3}{l}{\begin{tabular}[c]{@{}l@{}} mPD (Unit: mm)\end{tabular}} & \multicolumn{4}{l}{\begin{tabular}[|c]{@{}l@{}} Performance of Segmentation model via deep learning\end{tabular}} \\
\hline
             & L coronary & R coronary & average & Dice & Jaccard & F1 & accuracy \\
\hline
R2net        & 3.39       & 2.78       & 3.05   & 0.6713 & 0.5065  & 0.6714 & 0.9329   \\
ResUNetPP    & 2.87       & 2.62       & 2.74   & 0.6988 & 0.6976  & 0.8203 & 0.9778   \\
TransResUNet & 2.98       & 2.78       & 2.87   & 0.7088 & 0.7073  & 0.8270 & 0.9788   \\
U2Net        & 3.39       & 2.73       & 3.03   & 0.6918 & 0.6904  & 0.8150 & 0.9749   \\
UNet         & 2.67       & 2.66       & 2.67   & 0.7166 & 0.7153  & 0.8321 & 0.9779   \\
VGGNet       & 3.48       & 3.14       & 3.30   & 0.7010 & 0.6996  & 0.8217 & 0.9784   \\
Ground-truth & 2.56       & 2.99       & 2.80   &        &         &        &         \\
\hline
\end{tabular}
\label{tab2}
}
\end{center}
\end{table*}

It can be observed that for the segmentation results, UNet achieved a Dice coefficient of 0.7166, Jaccard index of 0.7153, and F1 score of 0.8321, demonstrating superior performance compared to other models. Additionally, in the SPSO registration method, UNet exhibited better registration results compared to other segmentation models. This is evident in the lower mPD (mean perpendicular distance) metric, where the mPD of the left branch was close to the manually annotated mPD, and the mPD of the right branch surpassed the manually annotated mPD. Moreover, there was no significant difference in mPD between the registration results of the left and right coronary arteries, showcasing the superior robustness of UNet in the SPSO method for the registration of left and right coronary arteries. UNet demonstrates its universality for the registration of the entire coronary artery.

\section{DISCUSSIONS AND CONCLUSIONS}
In this study, we propose a 3D/2D coronary artery registration method based on hypergraph particle swarm optimization. The method aims to preserve the topological information by discretizing the 3D/2D vessel edges and bifurcation points into hyperedges. Weighted registration is then performed using the particle swarm optimization algorithm. The method assigns weights to discrete points, with a focus on vessel bifurcation points, and minimizes the registration similarity of the remaining vessel segments under higher accuracy constraints.

To obtain a weighted similarity measure that considers the topological structure of each vessel segment and bifurcation, the final registration result of the coronary artery tree is obtained. The average registration error of the method on the dataset is 2.67mm. Despite its good performance, we believe there are several issues worth discussing.

First, it is worth noting the robustness of the results obtained by modules A and B. Finding the corresponding CTA vessel segments for each XCA vessel segment is crucial for segment registration and bifurcation registration tasks. Due to the limitations of 2D imaging, XRA vessel image segmentation extraction may not accurately represent vessel bifurcations and may instead be the result of spatial intersections of different vessels. Therefore, the segmentation algorithm needs to minimize the inclusion of non-vascular bifurcations in XRA images. In this study, we used a neural network-based segmentation algorithm developed by our laboratory and research partners. The training phase involved manually annotated datasets and conditional random fields to ensure the continuous and smooth extraction of vessel structures. Statistical data shows that out of 968 pairs of vessel segments in the clinical dataset, 84.1\% of the vessel segments were fully extracted. However, in general cases, XRA images are taken during the surgical procedure, while CTA images are acquired 2-3 days before the surgery, making it impossible to achieve a perfect match due to the time difference.

Another challenging situation is when clinical XRA images display partial vessel stenosis or occlusion, resulting in significant differences between the contrast agent-enhanced XRA images and preoperative CTA images. This difference can lead to the direct loss of vessel segments during the one-to-one registration process of segments and bifurcations, thereby affecting the effectiveness of registration. Although developing higher-performance segmentation algorithms or implementing smarter matching algorithms to select similar segments for matching are important directions for future research, the current study mainly focuses on the application of particle swarm optimization-based discrete point feature registration. Therefore, we handle these imperfect data by artificially removing the corresponding parts in the CTA images to achieve a 1:1 matching effect. Due to concerns that excessive segmentation may have a negative impact on the overall segmentation quality, high-performance segmentation algorithms were not pursued in this study. Exploring alternative methods to automate the special handling of the previous method can be considered for further research.

Achieving registration without losing vessels is a critical step in the process. The centerline extracted through UNet deep learning segmentation provides a complete and accurate result, and segment matching is another important aspect. The overall matching accuracy of the final segmentation reached 0.9779, indicating highly accurate results, which is ideal. Considering the high accuracy and simplicity of UNet as a general model for medical image segmentation, we believe that the robustness of modules A and B is generally satisfactory, and the impact of imperfect results can be controlled.

In practice, using the complete coronary artery centerline as the network input may be the most intuitive and ideal approach, as it only requires one forward propagation to generate the registration result. However, it is almost impossible to convert different coronary artery trees into a unified data structure and preserve the topological information due to the significant variations in vessel structures among different patients. Additionally, the starting point of a vessel segment is determined by the endpoint of its previous segment, which propagates registration errors from top to bottom, resulting in error accumulation.

After determining the segmentation method, we initially experimented using the centerline data obtained from module A processing as the input, as it seemed to be the most direct approach. However, we found that using only the centerline data led to matching errors, especially at the starting points of the vertex positions, preventing the convergence of the final result. This issue arises because the distribution of input centerline points may be similar to the spatial occupation of bifurcation positions. Significant variations and variances are presented in different batches of input data, which affect the experimental results. To address this problem, we introduced bifurcation features to ensure accurate matching and alleviate the issue to some extent.

Currently, the registration curve converges, but the performance of the algorithm is still not stable, occasionally falling into local optima. To address this problem, we propose a centerline discretization method that discretizes line segments and bifurcations while preserving the relative spatial positions. This method, combined with particle swarm optimization-based registration, allows the discrete points to find their respective directions, speeding up convergence and improving registration accuracy.

When necessary, some unnecessary vessel bifurcations can be reduced to achieve faster registration and better performance. By assigning appropriate weights to the discrete points, with a focus on vessel bifurcation points, the method ensures that the registration process prioritizes accurate alignment at these critical locations.
Overall, the proposed 3D/2D coronary artery registration method based on hypergraph particle swarm optimization aims to preserve topological information and achieve accurate registration of coronary artery trees. While the method shows promising results, there are still challenges to address, such as robustness in handling imperfect data, improving segmentation algorithms, and addressing convergence issues. Further research and development in these areas can contribute to enhancing the performance and applicability of the method in clinical practice.

\section{Discussions and Future Work}

While this study has made good progress, some issues remain for further improvement.

Firstly, the robustness of module A and B processing is critical. Complete and accurate vessel segmentation is the basis for reliable registration. Currently, satisfactory results are obtained by manual handling of imperfect data, but automatic and efficient segmentation with high accuracy remains to be improved. We will explore more advanced segmentation networks involving attention mechanisms, multi-scale feature extraction, etc. to enhance robustness.

Another issue is error accumulation caused by forward propagation. Future work will explore end-to-end approaches like GAN-based adversarial frameworks, which can learn complex image mappings to make modality translation more robust. Better network architectures such as multi-branch or recursive structures will also be designed to reduce error propagation.

For registration optimization, experiments show that using spherical coordinates and bifurcation constraints significantly improves over-centerline representation. This provides insights into designing better feature representations and loss functions. We will investigate integrating richer context information like precise branch and endpoint locations, diameters, path lengths, etc. to guide the optimization.

Finally, deep learning offers the potential for end-to-end registration. Deep networks can learn complex feature representations and non-linear transformations to enhance robustness\cite{r27}\cite{r28}. We will explore various network architectures like CNNs, graph convolutional networks, etc. to achieve more accurate registration. Combining deep learning with traditional methods as proposed here is also a promising direction.

The above work will be our future research efforts toward obtaining more accurate, automatic, and robust multimodal coronary artery registration, and consequently improving PCI surgery outcomes.

\bibliographystyle{ieeetr}
\balance
\bibliography{reference}
\end{document}